# Strain-modulated helimagnetism and emergent magnetic phase diagrams in highly crystalline MnP nanorod films

Richa Pokharel Madhogaria[1], Chang-Ming Hung[1], Baleeswaraiah Muchharla[1],

Anh Tuan Duong[2,*], Raja Das[2], Pham Thanh Huy[2], Sunglae Cho[3], Sarath Witanachchi[1],

Hariharan Srikanth[1,*], and Manh-Huong Phan[1,*]

[1] Department of Physics, University of South Florida, Tampa, Florida 33620, USA

[2] Phenikaa Research and Technology Institute, Phenikaa University, Yen Nghia, Ha-Dong District, Hanoi, 10000, Viet Nam

[3] Department of Physics and Energy Harvest-Storage Research Center, University of Ulsan, Ulsan 680-749, Republic of Korea

**ABSTRACT**

We explore strain-modulated helimagnetism in highly crystalline MnP nanorod films grown on Si(100) substrates using molecular beam epitaxy. The strained MnP film exhibits a paramagnetic to ferromagnetic (PM-FM) phase transition at $T_C \sim 279$ K, and the FM to helical phase transition at $T_N \sim 110$ K. The value of $T_N$ is greater than $T_N \sim 47$ K for the MnP single crystal, indicating strong strain-modulated helimagnetic states in the MnP nanorod film. The presence of significant thermal hysteresis in the helical phase indicates coexistence of competing magnetic interactions, leading to the first-order metamagnetic transition. Similar to its single crystal counterpart, an anisotropic magnetic effect is observed in the MnP film, which is independently confirmed by magnetic hysteresis loop and radio-frequency transverse susceptibility (TS) measurements. The evolution of screw (SCR) to CONE and FAN phase is precisely tracked from magnetization versus magnetic field/temperature measurements. The temperature dependence of the anisotropy fields, extracted from the TS spectra, yields further insight into the competing nature of the magnetic phases. Unfolding of the different helical phases at $T < 120$ K ($\sim T_N$) is analyzed by the temperature- and field-dependent magnetic entrsopy change. Based on these findings, the comprehensive magnetic phase diagrams of the MnP nanorod film are constructed for the first time for both the in-plane and out-of-plane magnetic field directions, revealing emergent strain/dimensionality-driven helical magnetic features that are absent in the magnetic phase diagram of the MnP single crystal.

*Corresponding authors: phanm@usf.edu (M.H.P.); tuan.duonganh@phenikaa-uni.edu.vn (D.A.T.); sharihar@usf.edu (H.S.)

## I. INTRODUCTION

Helimagnet is a form of antiferromagnetic structure where the spin twist along one direction results in the net magnetic moment to spatially rotate in a plane perpendicular to the propagation vector [1-3]. The helical spin configuration is usually the outcome of the antisymmetric Dzyaloshinskii-Moriya (DM) interaction in a non-centrosymmetric crystal structure [3-5]. DM interaction can also be realized in a centrosymmetric crystal with a broken space-inversion symmetry [1][6]. The other origins of the helical magnetic structure include competing magnetic interactions [7], with Ruderman-Kittel-Kasuya-Yoshida interaction mediated by the conduction electrons [8-9]. Owing to the robustness of the helicity against disturbances, helimagnets are of great importance for spintronics applications, especially for use in novel memory devices [1][10][11].

In addition to its excellent thermoelectric functionality, manganese phosphide (MnP), also known as a classical metallic helimagnet, has been the subject of intensive research since 1960's because of its exotic magnetic properties at low temperatures [12][13]. The modulation of spin structure caused by DM interaction gives rise to the non-trivial Hall effect in the FAN phase of MnP [14]. Recently, MnP has been established as an unconventional superconductor in which the application of pressure allows the tuning of antiferromagnetically to ferromagnetically mediated superconductivity [15]. A study by Jiang *et al*. has shown the control of spin helicity in MnP by using electric and magnetic fields [16]. MnP crystallizes into a centrosymmetric orthorhombic structure with the space group *Pbnm* ($a > b > c$), where each Mn atom is surrounded by six P atoms at four different Mn-P distances, giving rise to a distorted crystal structure compared to the NiAs type [12][13][17][18]. A detailed magnetic phase diagram evolves in bulk MnP comprising of the multiple metamagnetic phase transitions and Lifshitz critical behavior [18-20]. In the absence of an external magnetic field, bulk MnP undergoes a paramagnetic (PM) to ferromagnetic (FM) phase transition at $T_C$ = 292 K and then stabilizes into a helical (screw) phase below $T_N$ = 47 K. An additional transition at ~282 K, close to the $T_C$, owing to the spin reordering along the *b*-axis has been reported by Becerra *et al*. [21] and further confirmed through an AC susceptibility study performed by Yamazaki and coworkers [22]. A strong magnetocrystalline anisotropy behavior has been demonstrated in bulk MnP, with *a*, *b,* and *c*-axes being hard, intermediate, and easy magnetic axes, respectively [18-19]. The application of an external magnetic field emanates the screw (SCR) phase below 47 K to develop into the CONE and FAN phases, depending on the field orientation relative to the MnP crystal axis. In the SCR phase, the magnetic moments of MnP are confined into the *bc* plane, while for the CONE and FAN phases the net magnetic moment is along the direction of the applied magnetic field. Irrespective of the dimensions, the magnetic ground state of MnP is known to host multiple magnetic phase transitions. The MnP films grown on GaP substrates and the MnP nanocrystals embedded in the GaP epilayers showed the PM-FM phase transition at 291 K and 294 K, respectively, close to that reported for bulk MnP ($T_C$ = 292 K); however, there was a large increase in $T_N$, from 47 K for the bulk to 67 K for the film and 82 K for the nanocrystals [18]. The huge shift in $T_N$ has been attributed to the confinement of the nanostructures to a spatial region with a size comparable to the spread of one helix turn in the SCR phase, giving rise to the modified low temperature magnetic structures. The increase in $T_N$ (~100 K), for the FM-SCR phase transition, in the MnP film was also observed by Choi *et al*. and the presence of antiferromagnetic ordering up to 100 K was suggested to be associated with the strain induced due to the lattice mismatch between the epitaxial MnP film and the GaAs substrate [23]. However, the authors did not provide a detailed investigation on the strain and its influence on magnetism. On the contrary, a recent study on the 2D MnP single crystal has shown a different variation trend in $T_C$ and $T_N$; the $T_C$ increases to 303 K, while the $T_N$ decreases to 38 K [24]. The increase in $T_C$ is attributed to the dimensionality effect, but no comment on the $T_N$ has been made. Similarly, Monte Carlo simulations performed on MnP monolayers demonstrate the persistence of the FM ordering well above room temperature $T_C$ (495 K) with no signature of double-helical magnetic ordering at low temperatures [25]. Furthermore, the MnP nanorods size dependence of the magnetism showed a complete suppression of the $T_N$ for 20 nm long orthorhombic nanorods with their growth direction along the *b* axis [26]. The authors suggested that the absence of the AFM behavior in such a sample arose from the fact that the SCR phase was completely confined along the hard axis *a*.

Despite previous studies that revealed the competing magnetic phases in MnP in the forms of (bulk) single crystal, thin film [18][23] and nanocrystals/nanorods [18][26], a clear consensus about the dimensional effect on the magnetism of MnP has been yet to be reached. In the present study, we shed light on the dimensional and strain effects on the helimagnetism in a highly crystalline film consisting of MnP nanorods and have constructed, for the first time, the comprehensive magnetic phase diagrams for the in-plane and out-of-plane magnetic field directions, revealing emergent features that are absent in the magnetic phase diagram of its (bulk) single crystal counterpart. The paper is structured as follows: In Section III A, the structural and morphological characterizations of the MnP nanorod film are done via XRD, SEM, HRTEM, and SAED, which confirm the formation of single phase MnP. In Section III B, DC magnetic measurements, *M* vs. *T* and *M* vs. *H*, demonstrate the strain effect on the magnetism and the anisotropic magnetic behavior of the MnP nanorod film. In Section III C, the transverse susceptibility results further confirm the directional field dependence of the magnetization in the MnP film. The competing magnetic phases, identified via *M* vs. *T* and *M* vs. *H*, are also evident from magnetocaloric effect (MCE) measurements, as discussed in Section III D. A careful analysis of the temperature- and magnetic field-dependent magnetic entropy change ($\Delta S_M (T,\mu_0 H)$) unfolds the exotic magnetic phases in the MnP film as a result of the strain and cumulative effect of all three magnetic axes, unlike its single crystal counterpart, where *a*, *b* and *c* axes behave differently with respect to the external magnetic field.

## II. EXPERIMENT

MnP nanorod thin films were grown on Si (1 0 0) substrates by the molecular beam epitaxy method; the details of which have been reported elsewhere [17]. The films were grown at three different temperatures: 300 ºC, 400 ºC, and 500 ºC. The 400 ºC growth condition is optimal for the crystallinity and size of the nanorods, which is chosen for the present study. To further confirm the structure and the phase of the MnP film, high-resolution transmission electron microscopy (HRTEM) was done via TECNAI F20. Magnetic measurements were performed using a Quantum Design Physical Property Measurement System (PPMS) with a Vibrating Sample Magnetometer (VSM) option. Magnetization versus temperature was measured from $T = 3$ to 350 K for $H = 0.01–7$ T using zero-field cooled and field- cooled warming and cooling measurement protocols, and magnetization versus applied magnetic field (*M* vs. *H*) was measured from $H = 0$ to 5 T for temperatures ranging from $T = 5$ to 100 K. The diamagnetic signal from the substrate was detected in these measurements. The necessary procedure was therefore applied to subtract the diamagnetic signal from the measured *M* vs. *T* and *M* vs. *H* curves. Since a simple shift of the negative part of the magnetization in the *M* vs. *T* curve lead to the inappropriate increased magnetization value, we have used "arb. unit" for the *M* vs. *T* and *dM/dT* vs. *T* curves. Transverse susceptibility measurements were performed using a tunnel diode oscillator (TDO) with a resonant frequency of 12 MHz and sensitivity on the order of 10 Hz [27]. This measurement was done for $T = 20 – 300$

K with the applied field from $H = 0 - 50$ kOe. Magnetic entropy change was analyzed for the low temperature regime (3 K $< T <$ 120 K) from magnetization versus field. The data was collected using warming protocol for $\mu_0 H$ up to 5 T and with temperature steps of 3 K from $T = 3 - 40$ K, 2 K from $T = 40 - 120$ K. All the measurements were done for the in-plane and out-of-plane magnetic field directions with field and temperature sweep rates of 10 Oe/s and 5 K/min, respectively.

## III. RESULTS AND DISCUSSION

### A. Structural Characterization

The room temperature X-ray diffraction (XRD) pattern of the MnP film, as shown in Fig. 1(a), confirms the single orthorhombic phase with space group *Pbnm*. From the XRD peaks, the lattice parameters are determined to be $a = 5.839$ Å, $b = 3.164$ Å and $c = 5.301$ Å for the present MnP film, while those of the MnP single crystal are $a = 5.916$ Å, $b = 3.173$ Å, and $c = 5.260$ Å [17]. By evaluating the difference in the lattice parameters between the MnP single crystal and the MnP film, we find the presence of 1.30 % and 0.28 % *compressive* strain along the *a* and *b* axes and 0.78 % *tensile* strain along the *c*-axis in the MnP film, and a total of 0.81% volumetric change compared to the bulk MnP. The SEM image (refer to the inset of Fig. 1(a)) shows a uniform close packed distribution of the vertically grown nanorods on Si(100) substrate. The size of the nanorods is estimated to be ~100 x 50 nm (the length to diameter aspect ratio = 2) [17]. Figs. 1(b) and (c) show the cross-sectional HRTEM images of the MnP nanorod film. The cross-sectional shape of the rods is hexagonal (Fig. 1(b)), and the periodic atomic arrangement of the crystal with *d*-spacing between the crystal planes (2 1 0) to be ~2.58 Å (Fig. 1(c)) further confirms the formed MnP phase. In Fig. 1(d), the selected area electron diffraction (SAED) pattern shows a set of diffraction points that matches well with the plane identified from the HRTEM image, and provides additional points corresponding to the planes of the MnP single crystal. The detection of strain in the MnP film and its morphology (nanorods) makes us anticipate that these factors will have a significant impact on the magnetism of the film [18][26][28].

**B. Static Magnetization**

Figs. 2(a) and (b) show the magnetic field dependence of dc magnetization under zero-field-cooled (ZFC) protocol for $T = 3 - 300$ K and $H = 0.01 - 2$ T corresponding to the in-plane and out-of-plane field directions. With the increase of magnetic field from 0.01 T to 2 T, the development of different features below $T = 125$ K can be clearly seen for both the magnetic field directions (Fig. 2(a) and (b)) and is consistent up to the highest field measured (7 T) as shown in the insets of Fig. 2(a) and (b). The emergence of different attributes with the application of magnetic field in the *M* vs. *T* curve is more uniquely identified through $dM_{ZFC}/dT$ vs. *T* curves as can be viewed from Fig. 2(c) and (d), which illustrate the change in magnetization with respect to temperature for $T = 3 - 350$ K and $H = 0.01 - 2$ T for the in-plane and out-of-plane magnetic field orientations, respectively. The occurrence of peaks in the *dM/dT* vs. *T* curves can be understood as the sudden change in the spin ordering, leading to a significant change in the magnetization. With this understanding, the maxima and the minima in the *dM/dT* vs. *T* curves are assigned to the phase transitions that could possibly take place in the MnP film. The insets (i) of Figs. 2(c) and (d) illustrate the *M* vs. *T* and $dM_{ZFC}/dT$ vs. *T* curves for $\mu_0H = 0.1$ T to distinctly correlate the two magnetization curves. The minima in $dM_{ZFC}/dT$ vs. *T* for $\mu_0H = 0.1$ T at $T_C^{IP} = 278$ K represents the PM-FM phase transition for the in-plane field direction (inset (i) of Fig. 2(c)) and the positive peak ($T_{1\_IP} = 112$ K) shown by the black dashed line, can be associated with the FM-helical phase transition. Likewise, the negative and the non-negative peaks in the $dM_{ZFC}/dT$ vs. *T* curve at $\mu_0H = 0.1$ T for the out-of-plane field direction (inset (i) of Fig. 2(d)) can be recognized as the $T_C^{OP} = 280$ K and the $T_{1\_OP} = 108$ (the FM-helical phase transition temperature), respectively. The insets

(i) of Figs. 2(c) and (d) allow us to define the approximate $T_C$ and $T_N$ for the MnP nanorod film to be ~279 K and ~110 K, respectively. At $\mu_0H$ = 2 T, multiple peaks are seen in the $dM_{ZFC}/dT$ vs. $T$ curves for $T$ < 75 K (see inset (ii) of Fig. 2(c) and 2(d)) owing to the multiple features observed in the $M$ vs. $T$. In inset (ii) of Fig. 2(c), the peaks are marked by the dashed lines with red, green, and magenta, and assigned as $T_{2\_IP}$, $T_{3\_IP}$ and $T_{4\_IP}$, respectively [Fig. 2(c)]. The colors of the labels ($T_{1\_IP}$, $T_{2\_IP}$, $T_{3\_IP}$ and $T_{4\_IP}$) in Fig. 2(c) and the dashed lines in the inset (i) and (ii) of Fig. 2(c) are kept the same to show the one to one correspondence. Similarly, inset (ii) of Fig. 2(d) marks the various transitions by the colored dashed lines and are labeled as $T_{2\_OP}$ and $T_{3\_OP}$ [Fig. 2(d)] for the out-of-plane field direction (Note: the same colors are used for the dashed lines and the labels). The different magnetic transitions for both in-plane ($T_{1\_IP}$, $T_{2\_IP}$, $T_{3\_IP}$ and $T_{4\_IP}$) and out-of-plane ($T_{1\_OP}$, $T_{2\_OP}$ and $T_{3\_OP}$) field directions are discussed in detail below while constructing the magnetic phase diagrams for the MnP film.

Figs. 3(a) and (b) display the magnetization under field-cooled-cooling (FCC) and field-cooled-warming (FCW) protocols for $T$ = 3 − 300 K and $\mu_0H$ = 0.01− 2 T corresponding to the in-plane and out-of-plane field configurations. The presence of thermal hysteresis in the FCC and FCW of $M$ vs. $T$ curves is due to the supercooling or superheating phenomenon, leading to the discontinuous magnetic transitions. The thermal hysteresis in the $M$ vs. $T$ curves is usually considered as a signature of a first order phase transition (FOPT) [29]. The difference in the magnetic state of the system depending upon how the measurements have been taken can also be associated with the competitive nature of the magnetic interactions, which behave differently upon cooling or heating. The noted difference between the FCW and FCC $M$-$T$ curves at $\mu_0H$ = 0.1 − 7 T is shown in Fig. 3(c) and (d) for the in-plane and out-of-plane magnetic field directions, respectively. In Fig. 3(c), a significant hysteresis effect is seen below $T$ = 125 K for all the measured fields, indicating that the metamagnetic transitions (SCR to CONE and FAN) occurring at low temperatures ($T < 117$ K) in the MnP film are of FOPT type. Similarly, in Fig. 3(d), apart from the observed thermal hysteresis at $T < 125$ K, a noticeable difference in the FCC and FCW $M(T)$ curves is also observed for $T > 125$ K. To verify the presence of the hysteresis across the PM-FM transition, the FCC and FCW curves were remeasured at $\mu_0 H = 1$ T for a slower temperature sweep rate (2 K/min) and by stabilizing at each temperature. The thermal hysteresis obtained using different scan rates has been shown in the inset of Fig. 3(d). From the figure it can be seen that, lowering the sweep rate from 5 K/min to 2 K/min lowers the hysteresis effect around ~ 280 K, and is completely suppressed when the measurement is performed by stabilizing at each $T$ value. This is expected, as slow scan rate or steady temperature at each point gives more time for the sample to thermally stabilize. However, the significant hysteresis at low temperature remains persistent as a result of the competitive magnetic interactions.

The magnetic field dependence of magnetization at selected temperatures below 100 K [Fig. 4] shows a hysteresis loop, signifying that the low temperature spin ordering in the MnP film is not a canonical FM state. The butterfly-type shape of the curve, which is stressed at the origin indicates the presence of multiple magnetic phases in the MnP film. To understand the anomalous nature of the magnetic hysteresis curve, the virgin loop was scrutinized in detail. The insets (i) of Fig. 4(a) and (b) show the magnetic field dependence of differential susceptibility ($dM/dH$) for $T = 5$ K calculated using the virgin loops. The presence of the clear peak in the $dM/dH$ vs. $H$ curve suggests the sudden change of magnetization with the changing external magnetic field at a constant temperature; a metamagnetic transition, the feature that is expected in the case of our present system. The $H_{cr\_IP}$ and $H_{cr\_OP}$ are the critical fields required for the field-dependent transition to take place when the external magnetic field is applied parallel and perpendicular to the thin film, respectively. Furthermore, on comparing the *M* vs. *H* curves for the in-plane [Fig. 4(a)] and out-of-plane [Fig. 4(b)] magnetic field directions, we can conclude that the in-plane field configuration is more anisotropic than the out-of-plane one. The coercive field ($H_C$) vs. *T* curves shown in the insets (ii) of Fig. 4(a) and (b) validate the parallel field orientation to be more anisotropic than the perpendicular with respect to the film surface, as the value of $H_C$ for a given *T* is greater for in-plane compared to out-of-plane field direction. The presence of coercivity in the MnP film further confirms the low dimensionality of the grown sample, compared to the absence of any coercive fields in the case of polycrystalline or single crystal MnP [18][30]. In addition, the value of $H_C$ decreases smoothly for $T = 5 - 100$ K, increases at $T = 150$ K and further lowers with the rise in temperature beyond $T = 150$ K [inset (ii) Fig. 4 (a)]. Likewise, the value of $H_C$ decreases smoothly for $T = 5 - 100$ K, increases at $T = 100 - 200$ K and further lowers with the rise in temperature beyond $T = 200$ K [inset (ii) Fig. 4 (b)]. The temperature dependence of the coercive field shows a dissimilar trend as for the case of a pure FM, however, it reflects correspondence with the multiple magnetic phase transitions occurring in the MnP thin film. A similar behavior was also reported by de Andrés *et al*. where the variation of the coercive field with temperature was related to the *dM/dT* vs. *T* curve, and the magnetic phase transitions [18].

Figure 5 contrasts the phase diagrams between the MnP single crystal and the MnP nanorod film for $T < 120$ K. Figs. 5(a), (b) and (c) correspond to the phases for the hard, intermediate and easy axes, respectively, in the single crystal [18] and Figs. 5(d) and (e) represent the phase seen in the MnP film for the magnetic field orientations, in-plane and out-of-plane, respectively. For reference, the spin configurations of SCR, CONE and FAN are also illustrated in Fig. 5. In the case of the MnP single crystal, when the magnetic field is applied parallel to the *c*-axis for $T < 120$ K, only two known phases are observed; the FM phase where the spins are oriented parallel to the $c$-axis and the SCR phase for $T < 47$ K for low fields. However, for $\mu_0H > 0.26$ T, only FM // $c$-axis phase is present. For $\mu_0H$ // $b$-axis, although the scenario for low fields are similar ($T = 47$ K, $\mu_0H = 0.034$ T; FM – screw phase transition) to that of the $c$-axis, multiple metamagnetic phases come into the picture with the increase in $\mu_0H$. For $T \sim 1$ K, consecutive transitions from the SCR to FAN and FAN to the FM with spins parallel to the $b$-axis take place at $\mu_0H = 0.66$ T and 3.57 T, respectively, for $\mu_0H$ // b-axis. Similarly, in the case of the harder magnetic axis, $a$-axis, of the single crystal, the ground state is the SCR structure, which converts into the CONE phase for small fields and then to the FAN phase for $\mu_0H \geq 6$ T. When the system is warmed up, the FAN phase transition to the FM phase with spins parallel to the $c$ and $a$ axes for ($T, \mu_0H$) = (63 K , 4.32 T) and (75.7 K, 6.63 T), respectively.

The magnetic phase diagrams for the MnP nanorod film [Fig. 5(d) and (e)] are constructed based on the features seen in the $dM/dT$ vs. $T$ and $dM/dH$ vs. $H$ curves. In the present MnP film, the following two concept is considered while mapping out the phase evolution: the presence of strain caused by the lattice mismatch and the surface tension from the morphology of the nanorods in the film can play a significant role in the thermodynamic stability of the magnetic phases. Accordingly, different magnetically ordered phases are defined in Figs. 5(d) and (e). For the in-plane field direction, as described in the $dM/dT$ vs. $T$ plot [Fig. 2(c)], the FM-SCR phase transition takes place at $T_N^{IP} = 112$ K, while the PM-FM transition occurs at $T_C^{IP} = 278$ K. Similarly, the corresponding phase transitions occur at $T_N^{OP} = 108$ K and $T_C^{OP} = 280$ K [Fig. 2 (d)] when the external field is applied perpendicular to the film surface. According to Hirahara *et al*., the application of pressure on the $a$-axis of the MnP single crystal led to the enhancement of the AFM interactions and reduction of the FM one, resulting in the increased $T_N$ and decreased $T_C$ [30]. However, compression on the $c$-axis showed a reverse effect, with lower $T_N$ and higher $T_C$ values.

A study of the MnP nanoclusters embedded in GaP layers experiencing 0.5% volumetric strain compared to the bulk sample, shows a significant increase in $T_N$ (82 K) and almost no change in $T_C$ (294 K) [18]. Although the strain due to lattice mismatch was detected in the studied sample, the authors ascribed the combined trend of $T_N$ and $T_C$ to the size of the nanoclusters and surface effects. They also suggested that the magnetic interactions on the surface were affected by the strain and disorder (associated with increase of $T_N$), while the inside of the grain resembled the bulk feature (reflected by the unchanged $T_C$) [31]. In the present case, an opposite effect on the $T_N$ and $T_C$ is seen in the thin film, indicating stronger AFM interactions up to 110 K and comparatively, weaker FM interactions. This effect is similar to that of the bulk MnP, where the application of the strain increases the $T_N$ but decreases the $T_C$ [30]. Based on the earlier studies [18][26][30][31], the different variation trend of the transition temperatures for the in-plane and out-of-plane field geometry for the present MnP film indicates that the competitive magnetic interactions in the film are probed by the induced strain due to the change in the unit cell (0.81% change in the volume), the morphology of the film (nanorods) and the surface effects on the nanorods (length ~100 nm). However, the combination of all three effects reflect a different scenario, depending upon the direction of an external applied magnetic field.

*Magnetic phase diagram of the MnP film for the in-plane field orientation*: The temperatures, $T_{1\_IP}$, $T_{2\_IP}$, $T_{3\_IP}$ and $T_{4\_IP}$, as shown in Fig. 5(d) define the critical temperatures for various transitions taking place in the MnP film, and $H_{cr\_IP}$ is the critical field required to change the phase. Below $\mu_0 H = 0.1$ T, the FM state with the spins parallel to the plane of the film (FM2) transitions directly to the SCR phase, which is the magnetic ground state of the MnP film. For fields higher than 0.1 T, FM2 undergoes multiple magnetic phase transitions below and above $H_{cr\_IP}$. For $\mu_0 H > 0.1$ T, the FM2 ordering transitions to FM1 state at $T_{1\_IP}$, where FM1 has spins perpendicular to the film. In resemblance with the MnP single crystal, the FM1 and FM2 are two FM transitions corresponding to the easy and hard axes, out-of-plane and in-plane, respectively, as identified from the *M* vs. *H* measurements. Below $H_{cr\_IP}$ and for $\mu_0H > 0.1$ T, the FM1 orders into the CONE phase at $T_{2\_IP} \sim 58$ K. Similarly, the pure CONE phase transitions to the SCR phase for $\mu_0H = 0.1- 0.2$ T and the combined SCR and CONE phase for $\mu_0H = 0.5- 1.24$ T at $T_{3\_IP}$. The temperature $T_{4\_IP}$ guides the SCR+CONE phase to the pure SCR phase below $H_{cr\_IP}$. When the external magnetic field is greater than the critical fields, the magnetic ordering in the system takes a different path. For $\mu_0H = H_{cr\_IP} - 7$ T, it is assumed that the spin direction changes from the hard to easy axis, i.e. the FM2 state changes to the FM1 at $T_{2\_IP}$, which immediately transitions to the FAN phase at $T_{3\_IP}$. $T_{2\_IP}$ and $T_{3\_IP}$ are in close proximity to each other. Upon decreasing the temperature below 30 K, the FAN phase stabilizes into the combined CONE and FAN at $T_{4\_IP}$, for $\mu_0H > H_{cr\_IP}$. The CONE and FAN phases have the net magnetic moments in the direction of the externally applied magnetic field. Moreover, as mentioned above, since the axes are not well-defined in the case of the MnP nanorod film, the possibility of the mixed phases (SCR + CONE and CONE + FAN) is considered. Additionally, from the phase diagram of the MnP single crystal along the *a* and *b*-axes [Fig. 5(a) and (b)], it can be seen that the field required for the SCR phase to change into the FAN phase is comparatively higher than the critical fields required to convert the SCR into the CONE. With the understanding about the critical fields required for the SCR phase to convert into the CONE and FAN phase, the phase evolution of SCR below and above $H_{cr\_IP}$ has been assigned dominantly to CONE and FAN, respectively.

*Magnetic phase diagram of the MnP film for the out-of-plane field orientation:* Fig. 5(e) displays the magnetic phases present in the MnP film for $T < 120$ K and $\mu_0H = 0.01- 7$ T, when the external field is applied in the perpendicular direction. Unlike the parallel field orientation case, only single FM phase is present, justifying the out-of-plane field orientation as an easy magnetic axis. For $\mu_0 H < H_{cr\_OP}$ (critical fields responsible for the metamagnetic phase transition), the FM to SCR phase transition takes place at $T_{1\_OP}$. The persistence of the SCR phase is seen down to the lowest measured temperature ($T = 5$ K) below $H_{cr\_OP}$. However, when fields are greater than $H_{cr\_OP}$, multiple exotic phases are observed. For $\mu_0 H = H_{cr\_OP} - 7$ T, the FM-FAN phase transition occurs at $T_{2\_OP}$. The FAN phase finally orders into the combined CONE and FAN below $T_{3\_OP}$ and above $H_{cr\_OP}$. Similar to the in-plane field direction case, the net magnetic moments of the CONE and FAN phase are expected to be in the external field direction.

## C. Magnetocrystalline Anisotropy

The MnP single crystal is known to exhibit strong magnetocrystalline anisotropy, where the magnetic structure, CONE phase is confined only along the *a*-axis, FAN phase is seen along *a* and *b* axes and *c* being the easier magnetic axis with only SCR phase as the low temperature phase [12][13][18][19][20]. The influence of the preferred crystallographic orientations was also observed in the case of MnP nanocrystals embedded in GaP layers [18].

To explore the anisotropic nature of the MnP film, radio frequency transverse susceptibility (TS) measurements were performed using a custom built tunnel diode oscillator (TDO) probe [27][32]. The TS method provides more accurate information about how the magnetic anisotropy unfolds with temperature [33][34]. The change in resonant frequency ($\Delta f$) of the TDO circuit is a result of the change in inductance when the sample inside the circuit is magnetized. Therefore, $\Delta f$ is directly proportional to the change in TS ($\Delta \chi_T$) such that the quantity

$$\frac{\Delta \chi_T}{\chi_T}(\%) = \frac{|\chi_T(H) - \chi_T^{sat}|}{\chi_T^{sat}} x 100 \qquad (1)$$

can be measured as a function of $H_{DC}$, where $\chi_T^{sat}$ is the TS at the saturating or maximum field, $H_{sat}$. Peaks are expected in the TS scan at the positive and negative anisotropy fields, $\pm H_K$, and at the switching field, $-H_S$ [33][35].

TS scans were recorded for two different field orientations at different temperatures. Figs. 6(a) and (b) illustrate the TS results for the in-plane and out-of-plane field directions for $T = 20$ K and 70 K, respectively. The arrows in the graphs indicate the path of the sweeping magnetic field. The anisotropy fields ($\pm H_K$) are marked by using the dotted lines when the magnetic field changes from $+ 5$ T to $– 5$ T. The switching field is generally merged with one of these peaks [33][34][35]. Moreover, the application of a 5 T magnetic field is insufficient to saturate the system for both field geometries [Fig. 6(a) and (b)], which suggests that the feature observed in the TS spectra is the contribution from the spins pointing away from the direction of the external magnetic field. Therefore, for the in-plane field geometry, the TS probes the dynamics of the spins pointing out-of-plane and vice-versa.

The temperature dependence of $H_K$ for both the out-of-plane and in-plane spin configurations is shown in Figs. 6(c) and (d), respectively. As expected from the magnetic hysteresis data [Fig. 4(a)], the value of $H_K$ is higher for the in-plane [Fig. 6(d)], further validating the parallel field to be the harder magnetic axis. Furthermore, the $H_K$ vs. $T$ follows a similar trend as the $H_C$ vs. $T$ [insets (ii) of Fig. 4], revealing the existence of the competing magnetic phases in the MnP film. Upon decreasing the temperature below 250 K, the $H_K$ rises smoothly upto $T = 100$ K [Fig. 6(c)]. The reduction in $H_K$ with the increase in temperature ($T = 100 – 250$ K) is a typical behavior of a FM system, and this corresponds to the phase diagram [Fig. 5(e)], where for $T > 104$ K, the system is predominantly FM. Below 100 K, the presence of the competitive phases in the system [Fig. 5(e)] leads to the change in the $H_K$ characteristic. In the region $T = 50 – 100$ K, the abrupt drop in the anisotropy field is seen, while on further decreasing the temperature below 50 K, $H_K$ rises. The rise/drop in $H_K$ has been previously associated with the presence of the multiple magnetic orderings of the system [27]. In reference to the phase diagram [Fig. 5(e)], the approximate temperature regime, 50 − 100 K, hosts more than one phase transition involving AFM-AFM or AFM-FM spin re-ordering. Thus, the decrease in $H_K$ in the same temperature regime can be assigned to the presence of the competitive AFM and FM phases. Lastly, the low temperature (< 50 K ) behavior of $H_K$ is due to the predominant AFM spin ordering as seen from the phase diagram. Similarly, the three different temperature regimes, 250 − 300 K, 150 − 250 K and 20 − 150 K, in the $H_K$ vs. $T$ for the in-plane spin configuration [Fig. 6(d)] can be attributed to the FM2, FM1 and FM2, and AFM regions, respectively, with reference to the phase diagram shown in Fig. 5(d). The preferred spin direction and thermal fluctuations together play a major role to lower the energy barrier that the external field must overcome to align spins against their anisotropy axis, thus resulting in the temperature and spin dependent behavior of $H_K$. The correspondence of the temperature dependent $H_K$ with the magnetic phase diagram [Figs. 5(d) and (e)] signifies that the magnetic anisotropy plays an important role in the development and stabilization of the different magnetic phases in the MnP film, similar to the case of the bulk MnP.

### D. Magnetocaloric Effect and Magnetic Phase Diagrams

In addition to the understanding of the phase evolution using $M$ vs. $T$ and $M$ vs. $H$ data and TS studies, a detailed analysis of the static magnetic behavior across the $H$-$T$ phase diagram is conducted by utilizing the magnetocaloric effect (MCE). MCE is an effective tool used for probing phase coexistence and unraveling their temperature and field dependent characteristic features as the comprehensive magnetic phase diagrams of various magnetic systems [3][34][36][37][38][39]. Subjecting the magnetic sample to the change of external magnetic field at constant temperature, the magnitude of the change in isothermal magnetic entropy ($\Delta S_M$), can be numerically calculated by integrating the following Maxwell relation:

$$\left(\frac{\partial S_M(T,H)}{\partial H}\right)_T = \mu_0 \left(\frac{\partial M(T,H)}{\partial T}\right)_H \qquad (2)$$

which on integration between the applied field range yields the numerical value of $\Delta S_M$ as

$$\Delta S_M = \mu_0 \int_{H_i}^{H_f} \left(\frac{\partial M}{\partial T}\right) dH' \qquad (3)$$

The occurrence of maxima and minima in $\Delta S_M$ is the reflection of the slope change of the $M$ vs. $T$ curve at different fields (equation 3). The presence of peaks in the $\Delta S_M$ vs. $T$ curve represents a magnetic phase transition. In general, when there is a reduction of the magnetic contribution of the total entropy as the moments tend to align in the direction of the external field, suppressing thermal fluctuations, $\Delta S_M < 0$, i.e. $\Delta S_M$ vs. $T$ exhibits a minima [36-39]. On the contrary, when the external field induces spin disorder as the magnetic field forces the spins to align against their zero-field or ground state orientation, $\Delta S_M > 0$ is expected [36-39]. In addition, the spins are aligned with the external field direction only when the Zeeman energy overcomes the magnetic anisotropy energy, which further leads to non-negative change in the magnetic entropy [36-39]. In the following discussion, the magnetic entropy change is analyzed as functions of temperature and magnetic field across the $H$-$T$ phase diagram for the MnP film for two different field orientations.

From the isothermal $M$ vs. $H$ curves obtained in the region $T = 3 - 120$ K, $\Delta S_M$ is calculated for $\mu_0 \Delta H = 0 - 5$ T. Fig. 7 illustrates the $\Delta S_M(\mu_0 \Delta H, T)$ when the magnetic field was applied parallel to the thin film. Fig 7(a) shows $\Delta S_M(T)$ for $\mu_0 \Delta H = 0 - 0.8$ T. As the temperature is cooled below 120 K, a non-negative broad feature is observed at $T_1^{IP}$ [Fig. 7(a)], which corresponds to the $T_{1\_IP}$ seen in the $dM/dT$ vs $T$ curve. On further lowering the temperature, $\Delta S_M(T)$ exhibits a dominant minima at $T_2^{IP} \sim 57$ K and a maxima at $T_3^{IP} \sim 45$ K, which corroborate to the peaks, $T_{2\_IP}$ and $T_{3\_IP}$, of the $dM/dT$ vs $T$ curve. Likewise, a low temperature positive anomaly in $\Delta S_M$ at $T_4^{IP} \sim 28$ K is similar to our earlier finding, $T_{4\_IP}$. Irrespective of the field change, the $T_2^{IP}$, $T_3^{IP}$ and $T_4^{IP}$ remains persistent upto highest field measured ($\mu_0\Delta H = 5$ T), while the $T_1^{IP}$ is observed only up to 1 T, as shown in the inset of Fig. 7(a). Furthermore, the sharp increase in $\Delta S_M$ at $T_2^{IP}$ and $T_3^{IP}$ suggests the occurrence of the first order phase transitions at these temperatures, consistent with the thermal hysteresis observed in the MnP film [Fig. 3(c)]. Fig 7(b) and (c) demark the various magnetic phase transitions at $\mu_0\Delta H = 1$ T and 3 T, respectively. For $\mu_0\Delta H = 1$ T, FM2-FM1 transition takes place at $T_1^{IP}$, giving rise to $\Delta S_M > 0$, the FM1-CONE phase change occurs at $T_2^{IP}$ with a minima in $\Delta S_M$, CONE transforms to a mixed CONE+SCR state with positive values of $\Delta S_M$ at $T_3^{IP}$, and finally the mixed magnetic state (CONE+SCR) transforms into the SCR phase at $T_4^{IP}$. At high fields, $T_1^{IP}$ is suppressed [Fig. 7(c)], the negative peak in $\Delta S_M$ at $T_2^{IP}$ now corresponds to the FM2-FM1 phase transition. For $\mu_0\Delta H = 3$ T, $T_3^{IP}$ and $T_4^{IP}$ are associated with the FM1-FAN and FAN-FAN+CONE magnetic phase transitions, respectively. It should be noted that for a metamagnetic system, depending upon the change in the applied magnetic field, the same characteristic features in the $\Delta S_M$ vs. $T$ curve can signify different magnetic re-ordering.

The $\Delta S_M(T)$ for the out-of-plane field direction is shown in Fig. 8. The dominant attributes in the $\Delta S_M$ vs. $T$ curve for different external fields are marked in Fig. 8(a), (c) and (e), while the corresponding phase transitions are identified in Fig. 8(b), (d) and (f). The behavior of $\Delta S_M$ for $T = 3 - 120$ K and $\mu_0\Delta H = 0.02 - 0.1$ T is illustrated in Fig. 8 (a). The presence of a non-negative peak in $\Delta S_M$ at $T_1^{OP}$ resembles the feature $T_{1\_OP}$ seen in the $dM/dT$ vs $T$ curve [Fig. 2(d)] and represents the FM-SCR phase transition for $\mu_0\Delta H \leq 0.1$ T [Fig. 8(b)]. As $\mu_0\Delta H$ exceeds 0.1 T, a new development in $\Delta S_M$ at $T_3^{OP}$ is observed [Fig. 8(c)]. The noted $\Delta S_M > 0$ at $T_3^{OP}$ resembles the $T_{3\_OP}$ from $M$ vs. $T$ measurements. However, $T_{3\_OP}$ in $dM/dT$ vs $T$ is seen for $\mu_0 H \geq 2$ T while MCE reveals its presence for $\mu_0 \Delta H \geq 0.2$ T relevant to two different phase transitions: 2 T $\geq \mu_0 \Delta H \geq$ 0.2 T and $\mu_0 H \geq 2$ T. For $\mu_0 \Delta H = 1$ T [Fig. 8(d)], unlike for low field change ($\mu_0 \Delta H \leq 0.1$ T), $T_1^{OP}$ corresponds to the FM-CONE and the subsequent $T_3^{OP}$ (2 T $\geq \mu_0 \Delta H \geq 0.2$ T) is the critical temperature for the CONE-SCR phase transition. For $\mu_0 \Delta H > 1$ T, a minima in $\Delta S_M$ evolves between $T_1^{OP}$ and $T_3^{OP}$ and is marked as $T_2^{OP}$ [Fig. 8(e)]. With the suppression of $T_1^{OP}$ for $\mu_0 \Delta H > 2$ T, the field evolution of $T_2^{OP}$ becomes more prominent, as can be seen from the inset of Fig. 8(e). The $T_2^{OP}$ is similar to the $T_{2\_OP}$ [Fig. 2(d)], nevertheless, the critical field for $T_2^{OP}$ from $\Delta S_M$ vs. $T$ is obtained to be 1.4 T, which is lower than the field (2 T) at which $T_{2\_OP}$ is noted in the $dM/dT$ vs $T$. As shown in Fig. 8(f), for $\mu_0 \Delta H = 2$ T, $T_2^{OP}$ marks the FM-FAN phase transition and dissimilar to the low fields, a FAN phase transforms into the mixed FAN+CONE at $T_3^{OP}$.

The $H$-$T$ phase diagrams [Fig. 9] for the in-plane and out-of-plane field directions are constructed based on the temperature and field dependences of $\Delta S_M$. The unfolding of FM and AFM phases as functions of temperature and field can be well illustrated from $\Delta S_M (\mu_0 \Delta H, T)$. The cool and warm colors in the figures signify the responses of the system to the external magnetic field, resulting in $\Delta S_M < 0$ and $\Delta S_M > 0$, respectively. The characteristic features noted in $\Delta S_M$ vs. $T$; $T_1^{IP}$, $T_2^{IP}$, $T_3^{IP}$ and $T_4^{IP}$ for the in-plane curves [Fig. 9(a)] and $T_1^{OP}$, $T_2^{OP}$ and $T_3^{OP}$ for the out-of-plane curves [Fig. 9(b)] are represented by line and symbol plots. Additionally, the maxima in the $dM/dH$ vs. $H$ from the MCE data is also shown. The $H_{cr}^{IP}$ and $H_{cr}^{OP}$ sketched by using star+line are the peaks in the differential susceptibility for the in-plane and out-of-plane, respectively. The phase evolution revealed by $\Delta S_M (\mu_0 \Delta H, T)$ further validates our earlier finding of the magnetic phases using the $M$ vs. $T$ and $M$ vs. $\mu_0 H$ [Fig. 5] data. A one to one correspondence between the in-plane field direction phase diagram obtained from $M$ vs. $T$ /$M$ vs. $\mu_0H$ and $\Delta S_M$ ($\mu_0\Delta H,T$) can be established. However, for the out-of-plane field direction, $\Delta S_M$ ($\mu_0\Delta H,T$) adds more features that are not seen in the $M$ ($T,\mu_0H$) curves.

Conventionally, a negative peak in $\Delta S_M$ is assigned to the PM-FM phase transition and $\Delta S_M > 0$ indicates an AFM ordering in the system [36-39]. However, this definition cannot be implemented for a system with multiple magnetic orderings, like MnP. The presence of any peaks in $\Delta S_M$ ($\mu_0\Delta H,T$) curve signifies the change in the magnetic ordering and depending upon whether the external field favors or disfavors the ground state spin configuration, a minima or a maxima is observed in $\Delta S_M$ ($\mu_0\Delta H,T$). For instance, the FM2-FM1 transition at $\mu_0\Delta H = 1$ T for the in-plane field orientation [Fig. 7(b)], has $\Delta S_M > 0$, which is expected as the spins are parallel to the easy axis in the case of FM1, while FM2 has spins pointed along the hard axis, hence in the process of transforming the spin direction from the easy to hard axis, spin disorder is created resulting in the positive $\Delta S_M$. Nevertheless, for the same transition at $\mu_0\Delta H = 3$ T [Fig. 7(c)], a minima in $\Delta S_M$ is observed. Here it should be noted that the external field is greater than the anisotropy field, $H_K \sim$ 2.5 T at $T = 20$ K [Fig. 6(d)], therefore, the Zeeman energy overcomes the magnetic anisotropy energy, giving rise to the spin-flip transition, and consequently $\Delta S_M < 0$.

From the magnetic phase diagram [Fig. 9], one can see that the MnP nanorod film exhibits features thar are different from the MnP single crystal. The mixed phases (SCR+CONE and CONE+FAN) are some of the distinct characteristics of the MnP nanorod film. The critical fields required for the conversion of the SCR phase to the CONE or FAN ($H_{cr}^{IP}$ and $H_{cr}^{OP}$) are higher in the case of the film, indicating the low temperature stability and robust nature of the SCR phase. Furthermore, the persistence of the CONE and FAN phases for $T < = T_3^{IP}$ (in-plane) and $T < =$ $T_2^{OP}$ (out-of-plane), upto the highest measured field, $\mu_0\Delta H = 5$ T, suggest that the presence of the strain and the combined effects of *a,b* and *c*-axes along with lower thermal fluctuations favor the AFM state.

## IV. CONCLUSION

A comprehensive magnetic study has been performed on the MnP nanorod thin film for the in-plane and out-of-plane field orientations. The presence of the compressive strain along the *ab*-plane of the grown film drastically affects the transition temperatures ($T_N$ and $T_C$) of the sample, stabilizing the AFM order upto higher temperature. The field dependent magnetization results suggest the anisotropic nature of the sample. The system has an easy magnetic axis when the field is applied perpendicular to the plane of the film and a hard axis when the field is parallel. The anisotropic behavior was further validated by the TS measurements, which establish the role of the magnetocrystalline anisotropy in the stabilization of the multiple magnetic phases in the MnP film. The co-existence of the competing phases was further demonstrated by the MCE study. The comprehensive magnetic phase diagrams for the in-plane and out-of-plane field directions have been constructed for the MnP nanorod film to illustrate the co-existing magnetic phases and unfold the strain/dimensionality-influenced features that are absent in its single crystal counterpart.

**ACKNOWLEDGMENTS**

Research at the University of South Florida was supported by the U.S. Department of Energy, Office of Basic Energy Sciences, Division of Materials Sciences and Engineering under Award No. DE-FG02-07ER46438. MHP, PTH and DAT acknowledge support from the VISCOSTONE USA under Grant No. 1253113200.

**FIGURE CAPTIONS**

**FIG. 1** Structural characterization of the MnP nanorod film. (a) XRD pattern, (b) and (c) the cross-sectional HRTEM images of the MnP film and (d) the SAED image of the film. Inset (a) is an SEM image of the vertically grown MnP nanorods at 400 °C on Si(1 0 0) substrate.

**FIG. 2** Temperature dependence of the magnetization, $M$ vs. $T$, under zero field-cooled (ZFC) protocol for $\mu_0 H = 0.01 - 2$ T. (a) the external field was applied parallel to the plane of the film (*in-plane*), (b) the external field was applied perpendicular to the plane of the film (*out-of-plane*). $dM_{ZFC}/dT$ vs. $T$ indicating different transition temperatures using dashed lines. For *in-plane*: $T_{1\_IP}$, $T_{2\_IP}$, $T_{3\_IP}$ and $T_{4\_IP}$ (c) and for *out-of-plane*: $T_{1\_OP}$, $T_{2\_OP}$ and $T_{3\_OP}$ (d). One to one correspondence between $M$ vs. $T$ and peaks observed in $dM_{ZFC}/dT$ vs. $T$ for $\mu_0 H = 0.1$ T and 2 T is shown in insets (i) and (ii) of (c) and (d), respectively.

**FIG. 3** Temperature dependence of the magnetization, $M$ vs. $T$, under field-cooled-warming (FCW) and field-cooled-cooling (FCC) protocols for $\mu_0 H = 0.01 - 1$ T. (a) for *in-plane* and (b) for *out-of-plane*. The difference in the magnetization ($|M_{FCC} - M_{FCW}|$) with respect to temperature is shown for (c) *in-plane* and (d) *out-of-plane*. Thermal hysteresis for different temperature scans at $\mu_0 H = 1$ T for out-of-field geometry is shown in the inset of (d).

**FIG. 4** Magnetization versus magnetic field, $M$ vs. $H$, at different temperatures: (a) *in-plane* and (b) *out-of-plane*. Field dependence of differential susceptibility, $dM/dH$ vs. $H$ at $T = 5$ K, inset (i) and the temperature-dependent coercive field ($H_c$), inset (ii), of (a) and (b), respectively.

**FIG. 5** Magnetic phase diagrams of the MnP single crystal for three different field geometries at $T \leq 120$ K (adapted from Reference [18]), (a) H // *a*-axis, (b) H // *b*-axis and (c) H // *c*-axis, and the magnetic phase digrams of the MnP nanorod film are constructed based on the features seen

in the *dM/dT* vs. *T* and *dM/dH* vs. *H* curves for $\mu_0 H = 0.01 - 7$ T and $T \leq 120$ K for two different field configurations, (d) *in-plane* and (e) *out-of-plane*.

**FIG. 6** Temperature dependence of effective magnetic anisotropy field of the MnP film using transverse susceptibility (TS). Bipolar TS scans as a function of applied magnetic field ($\mu_0 H = -5 - +5$ T), (a) at $T = 20$ K and (b) at $T = 70$ K showing *out-of-plane* and *in-plane* anisotropy, respectively. The peaks in TS denote the effective magnetic anisotropy fields, $\pm H_K$. (c) and (d) show the temperature dependence of $H_K$ for *out-of-plane* and *in-plane*, respectively.

**FIG. 7** The magnetic entropy change ($\Delta S_M$) for the external applied field parallel to the plane of the film (a) as a function of temperature for low fields, $\mu_0 \Delta H = 0.2 - 0.8$ T, and $T = 3 - 120$ K. The different features in the curves are marked by the black dashed lines and represented as $T_1^{IP}$, $T_2^{IP}$, $T_3^{IP}$ and $T_4^{IP}$. The inset shows the $\Delta S_M$ vs. *T* for high magnetic fields, $\mu_0 \Delta H = 1 - 5$ T, and $T = 3 - 80$ K. (b) and (c) identify the different possible phase transitions taking place in the film reflected by the characterictic features in the $\Delta S_M$ vs. *T* curves for two field vlaues, $\mu_0 \Delta H = 1$ T and $\mu_0 \Delta H = 3$ T, respectively.

**FIG. 8** Temperature-dependent $\Delta S_M$ for the external applied field perpendicular to the plane of the film for different magnetic fields and $T = 3 - 120$ K is shown. The development of the peaks are noted by the dashed lines and symbolized as $T_1^{OP}$, $T_2^{OP}$ and $T_3^{OP}$. (a), (c) and (e) for $\mu_0 \Delta H = 0.02 - 0.1$ T, $0.2 - 1$ T and $1.2 - 2$ T, respectively. Inset of (e) shows the $\Delta S_M$ vs. *T* for high magnetic fields, $\mu_0 \Delta H = 3 - 5$ T. The different possible phase transitions taking place in the film are marked in the $\Delta S_M$ vs. *T* curves for three field vlaues, $\mu_0 \Delta H = 0.1$ T (b), $\mu_0 \Delta H = 1$ T (d) and $\mu_0 \Delta H = 2$ T (f).

**FIG. 9** *H-T* phase diagrams of the MnP nanorod film constructed using the magnetic entropy data for two different field configurations, (a) *in-plane* and (b) *out-of-plane*. The surface plots show $\Delta S_M$ over the full range of temperatures and changes in magnetic field studied. The peaks observed in $\Delta S_M$ vs. *T* ($T_1^{IP}$, $T_2^{IP}$, $T_3^{IP}$ and $T_4^{IP}$ for the *in-plane*) and ($T_1^{OP}$, $T_2^{OP}$ and $T_3^{OP}$ for the *out-of-plane*) curves are shown by the line and symbol plots. The critical fields, $H_{cr}^{IP}$ and $H_{cr}^{OP}$, sketched by using star+line are the maxima in the *dM/dH* vs. *H* curves for the in-plane and out-of-plane, respectively. The different competitive magnetic phases at respective field and temperature are labeled.

**FIG. 1**

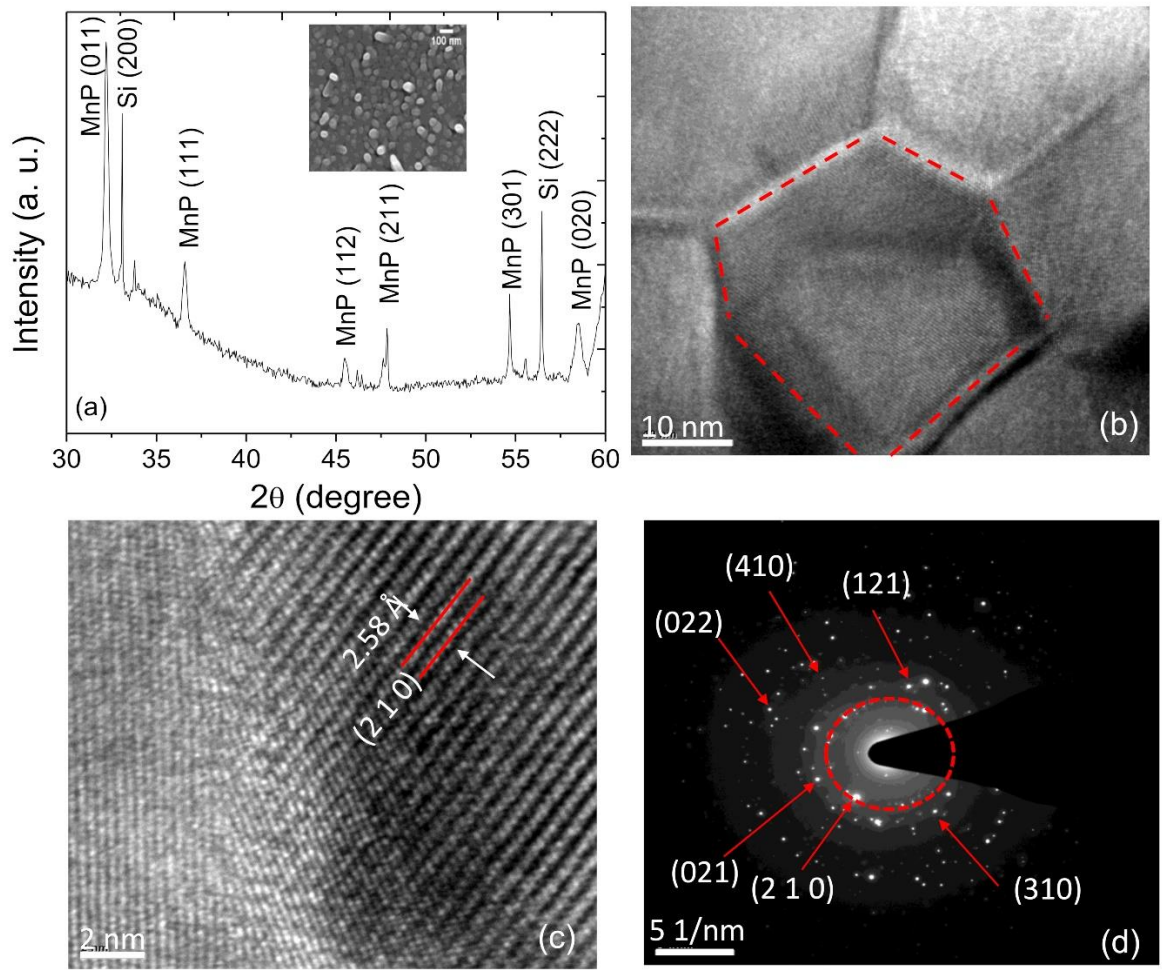

**FIG. 2**

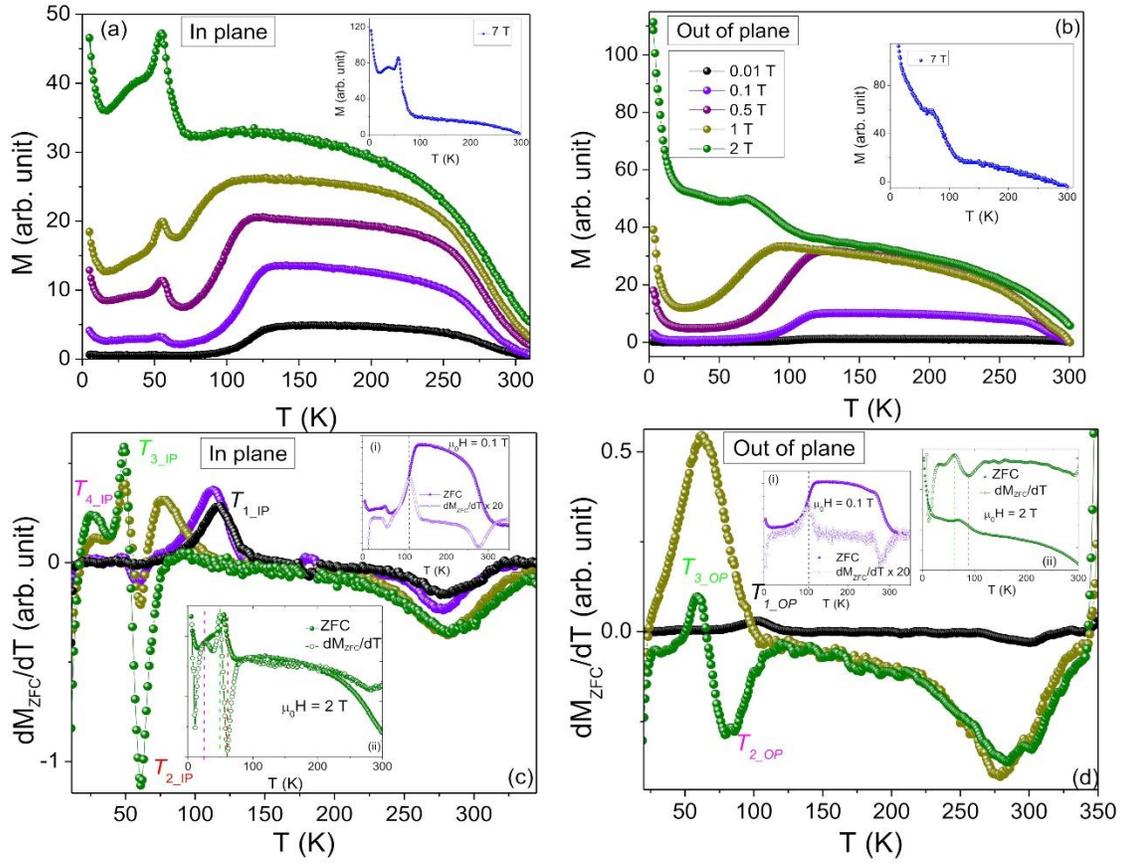

**FIG. 3**

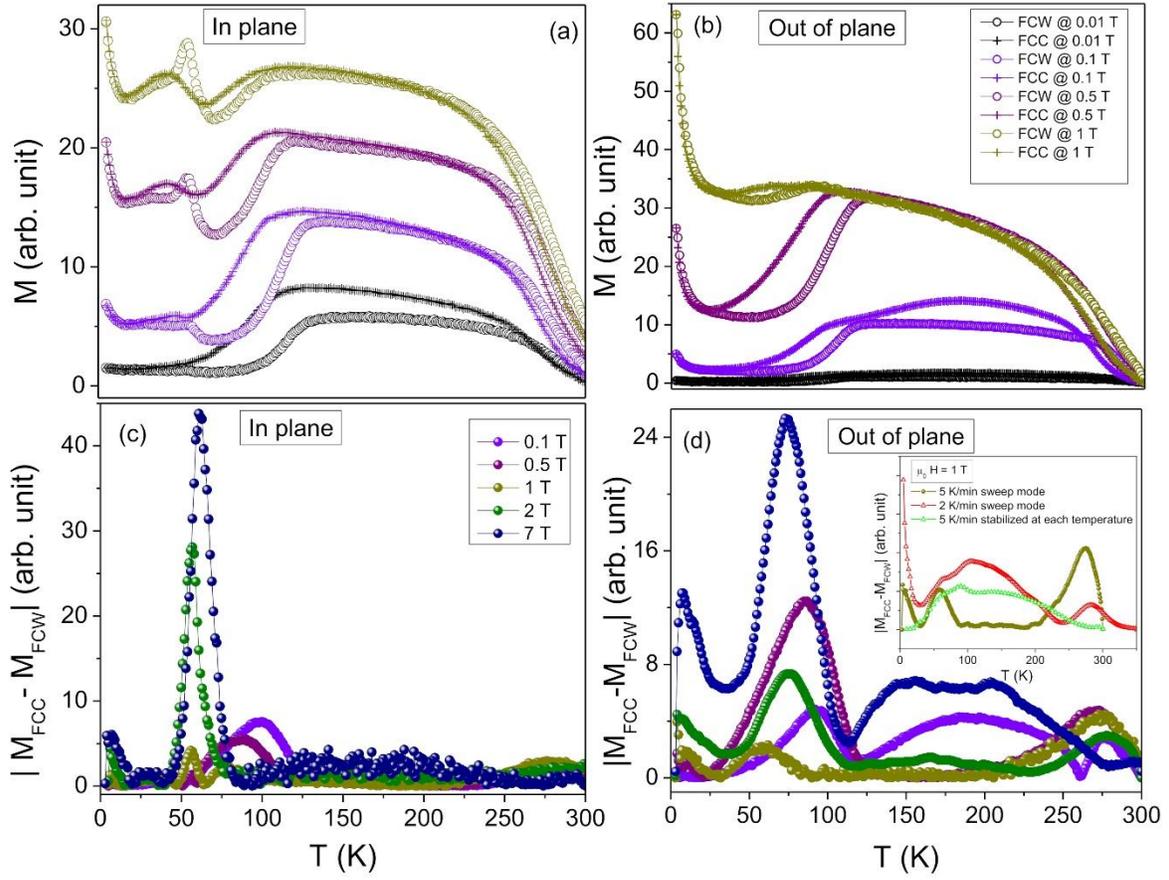

**FIG. 4**

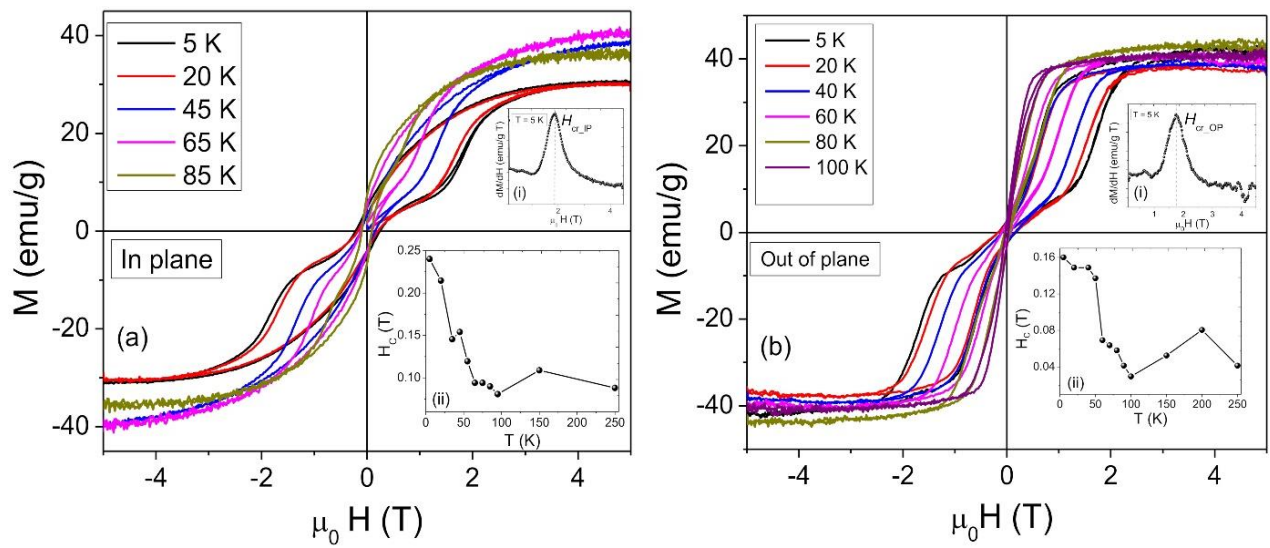

**FIG. 5**

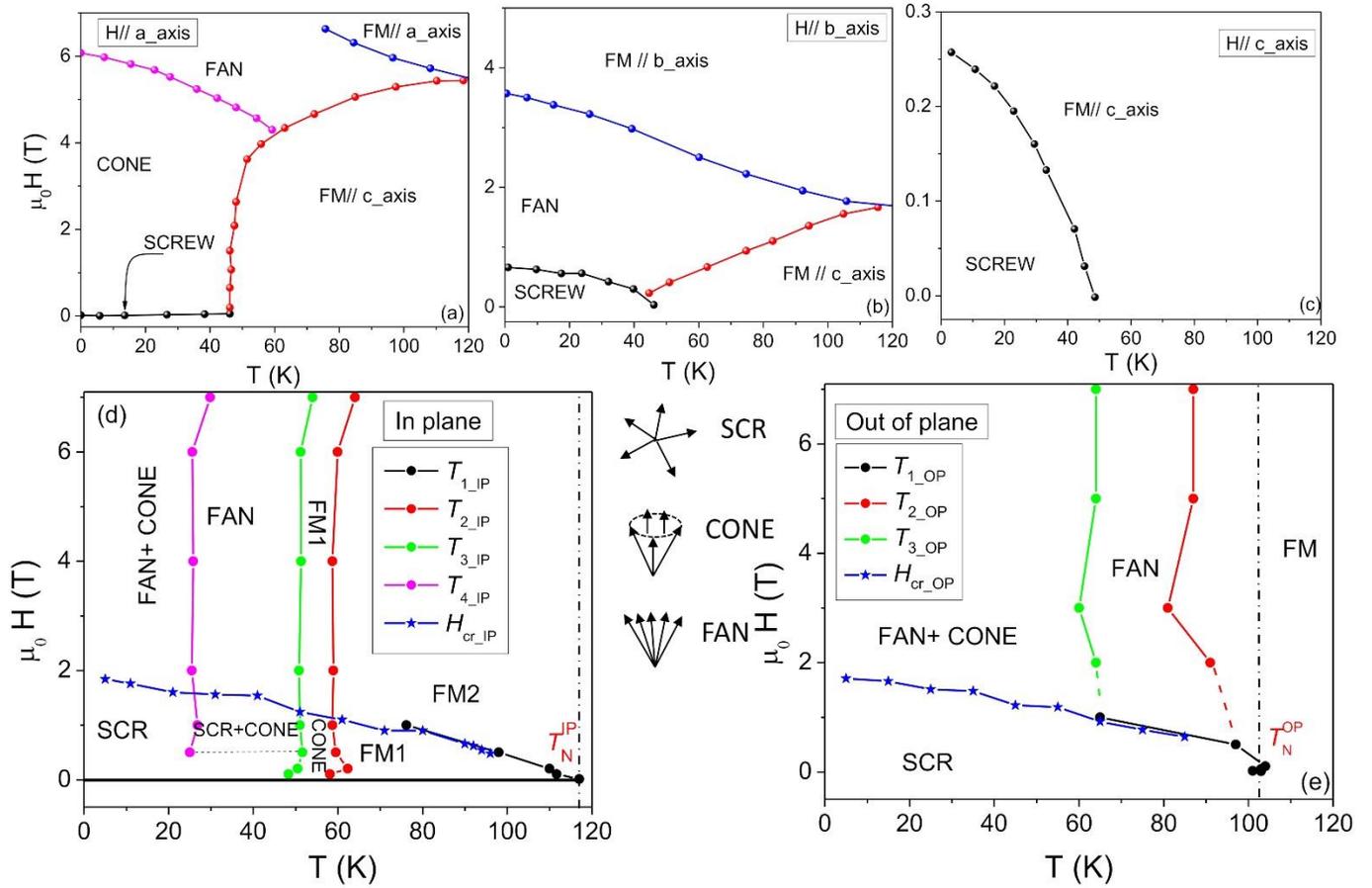

**FIG. 6**

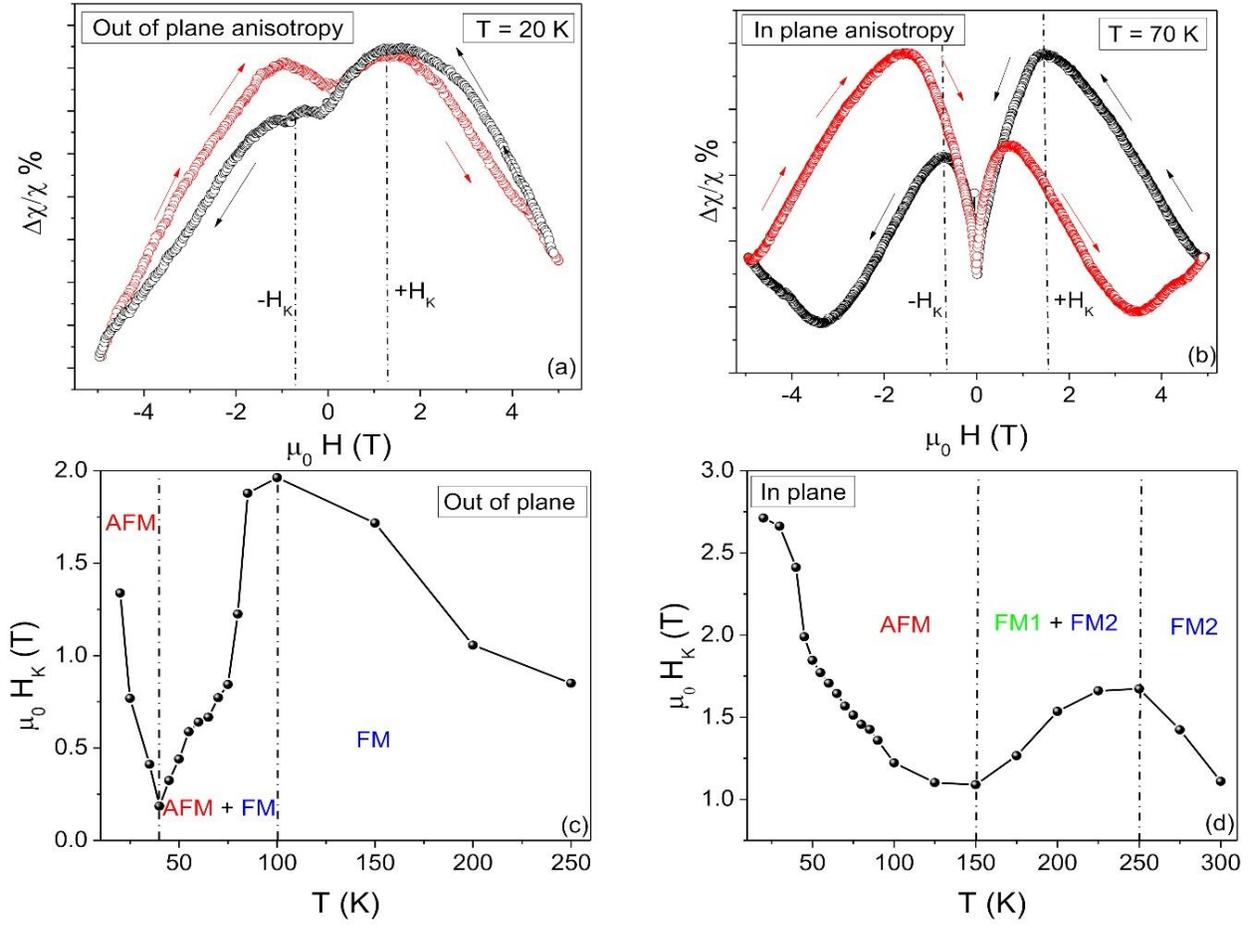

**FIG. 7**

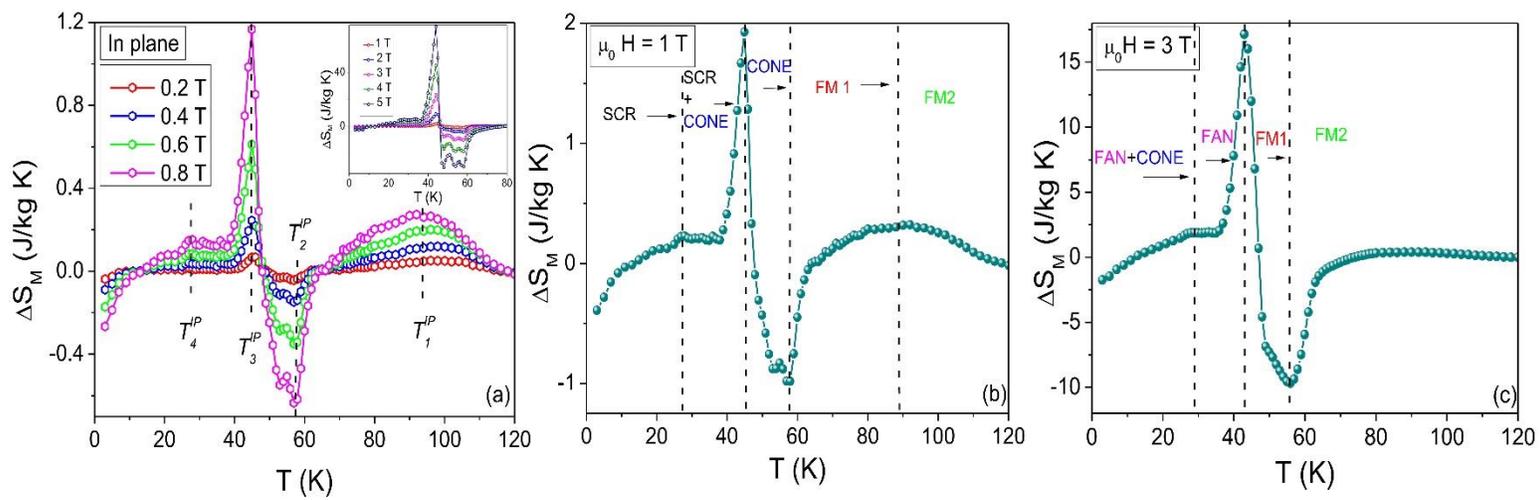

**FIG. 8**

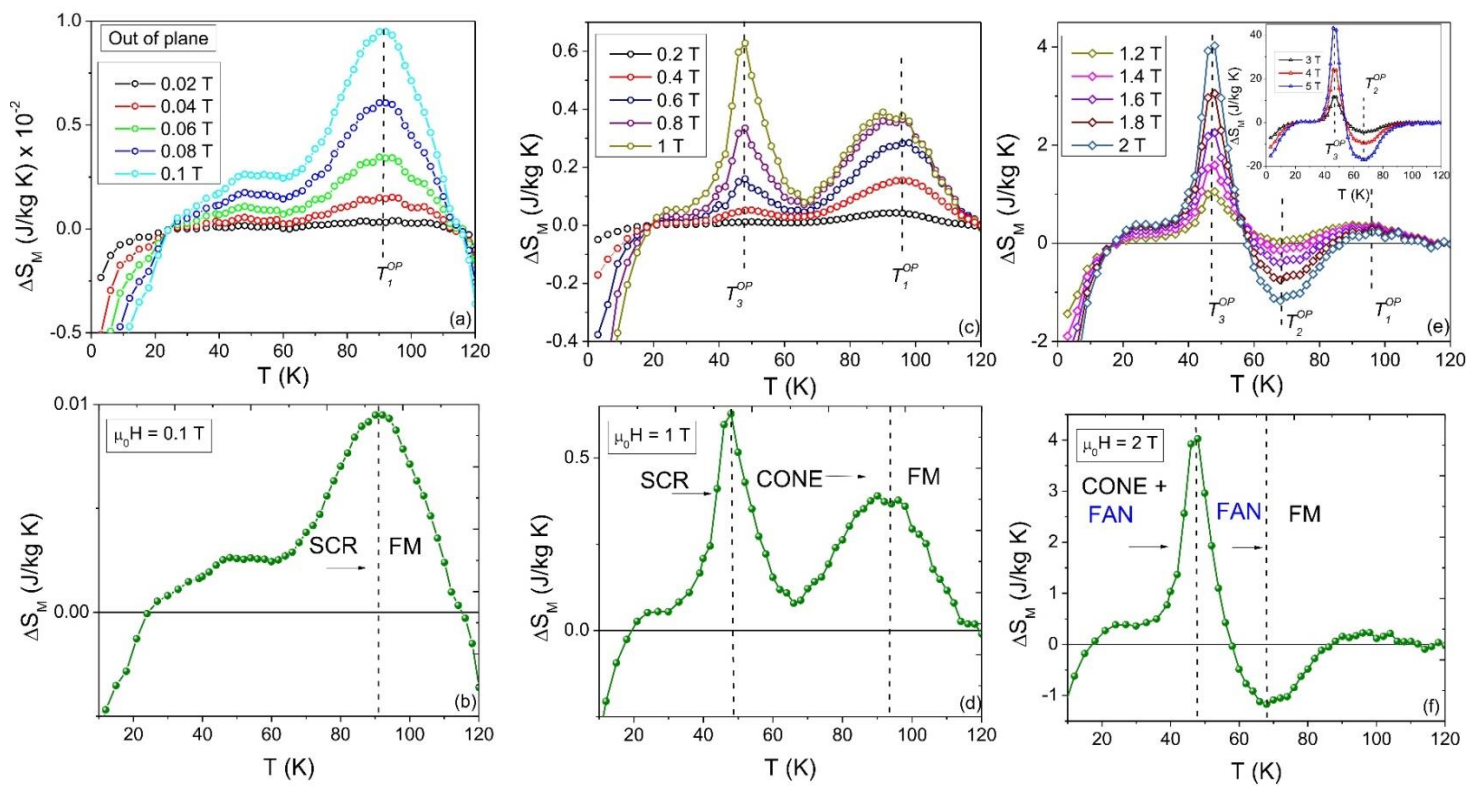

**FIG. 9**

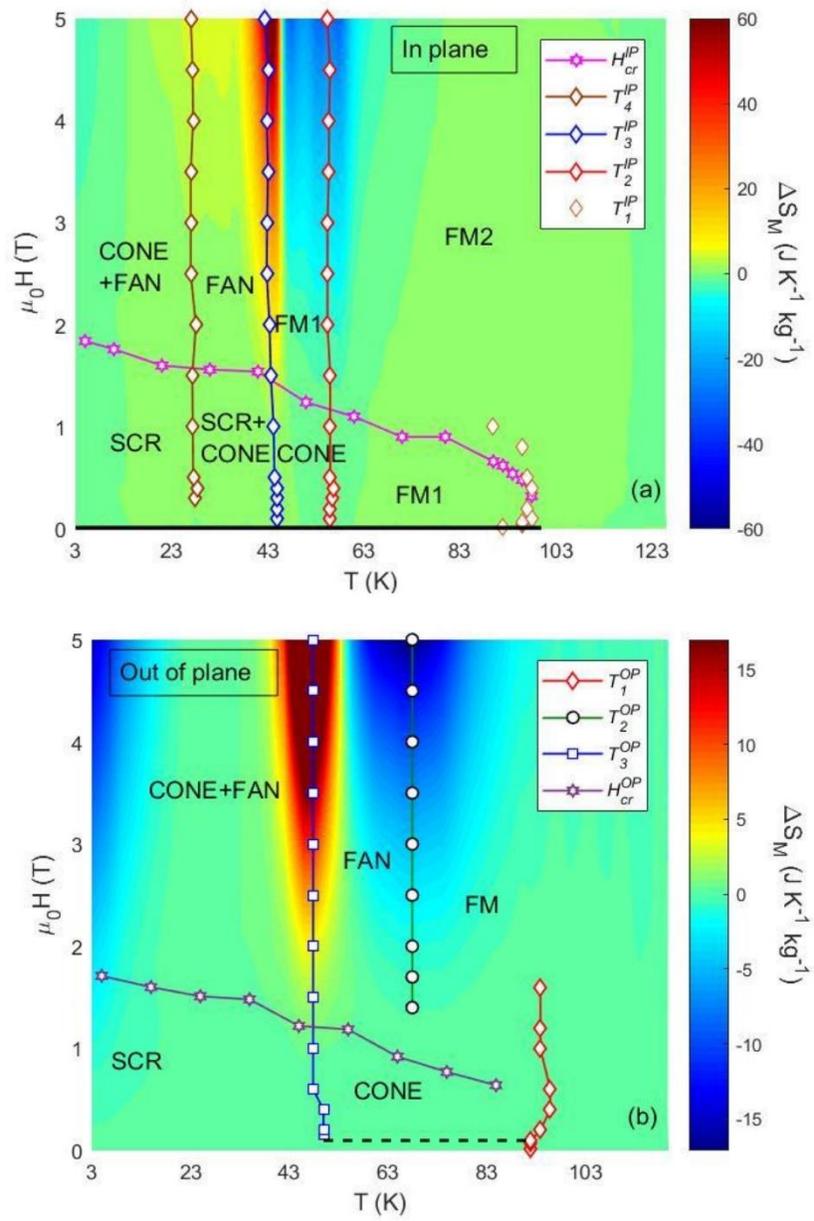